\documentclass[aps,prc,preprint,amsmath,showpacs,superscriptaddress,nofootinbib]{revtex4-1}
\usepackage{graphicx,epsfig,psfrag}

\begin{document}

\title{Role of Landau-Rabi quantization of electron motion on the crust of magnetars within the nuclear energy density functional theory}

\author{Y. D. Mutafchieva}
\affiliation{Institute for Nuclear Research and Nuclear Energy, Bulgarian Academy of Sciences, 72 Tsarigradsko Chaussee, 1784 Sofia, Bulgaria}
\author{N. Chamel}
\affiliation{Institute of Astronomy and Astrophysics, Universit\'e Libre de Bruxelles, CP 226, Boulevard du Triomphe, B-1050 Brussels, Belgium}
\author{Zh. K. Stoyanov}
\affiliation{Institute for Nuclear Research and Nuclear Energy, Bulgarian Academy of Sciences, 72 Tsarigradsko Chaussee, 1784 Sofia, Bulgaria}
\author{J.M. Pearson}
\affiliation{D\'ept. de Physique, Universit\'e de Montr\'eal, Montr\'eal (Qu\'ebec), H3C 3J7 Canada}
\author{L. M. Mihailov}
\affiliation{Institute of Solid State Physics, Bulgarian Academy of Sciences, 72 Tsarigradsko Chaussee, 1784 Sofia, Bulgaria}

\begin{abstract}
Magnetic fields of order $10^{15}$~G have been measured at the surface of some neutron stars, and much stronger magnetic fields are 
expected to be present in the solid region beneath the surface. The effects of the magnetic field on the equation of state and on the 
composition of the crust due to Landau-Rabi quantization of electron motion are studied. Both the outer and inner crustal regions are described 
in a unified and consistent way within the nuclear-energy density functional theory. 
\end{abstract}


\maketitle

\section{Introduction}
\label{intro}

The catastrophic gravitational-core collapse of massive stars during supernova explosions can produce so called \emph{magnetars}, i.e. 
neutron stars endowed with extremely high magnetic fields exceeding $10^{14}$~G~\cite{td92}. This huge magnetic energy is thought to 
power soft-gamma ray repeaters (SGRs) and anomalous X-ray pulsars (AXPs), as evidenced by measurements of surface magnetic fields 
up to a few times $10^{15}$~G (see, e.g. Ref.~\cite{kaspi17} for a review). 
Numerical simulations have shown that the internal magnetic field of a neutron 
star could potentially reach $\sim 10^{18}$~G (see, e.g. Refs.~\cite{kiuchi2008,frieben2012,pili17,chatterjee2015} and references therein).  	

The effects of a high magnetic field on a magnetar are most prominent in the outer crust, which consists of a succession of different strata made of atomic nuclei with proton number $Z$ and mass number $A$ in a charge-compensating electron background. We have previously shown that the composition and the equation of state in this region can be drastically modified due to Landau-Rabi quantization of electron motion~\cite{chapav12,bjp2013,chasto15,chamut17} (see also Ref.~\cite{blacha18} for 
a general review). 
The magnetic condensation of the surface layers~\cite{ruderman71, lai01} leads to a strong stiffening of the equation of state. Introducing the characteristic magnetic field strength 
\begin{equation}
\label{eq:Brel}
B_{\rm rel}=\frac{m_e^2 c^3}{e\hbar}\simeq 4.41\times 10^{13}\, \rm G\,
\end{equation}
(with $m_e$ the electron mass, $c$ the speed of light, $e$ the elementary electric charge, and $\hbar$ the Planck-Dirac constant), 
the increase of the mean baryon number density $\bar n$ with the pressure $P$ is approximately given by~\cite{chapav12}
\begin{equation}\label{magnetic_cond}
\bar n\approx \bar n_s \left(1+\sqrt{\frac{P}{P_0}}\right)\, , 
\end{equation}
where $\bar n_s\simeq 2.50\times 10^{-10} B_\star^{6/5}$ is the density at the stellar surface (made of iron), $P_0\simeq 1.82\times 10^{-11} Y_p^2 B_\star^{7/5}$ MeV~fm$^{-3}$, $Y_p=Z/A$ is the proton  fraction, and $B_\star=B/B_{\rm rel}$. 
With increasing magnetic field strength, the maximum density $\bar n_{\rm max}$ at which a nuclide is present exhibits typical quantum oscillations as electrons fill fewer and fewer Landau-Rabi levels~\cite{chamut17}. If electrons are confined to the lowest level, $\bar n_{\rm max}$ increases linearly with $B_\star$ (ignoring the lattice correction). The magnetic field is strongly quantizing in all regions of the outer crust if $B_\star\gtrsim 1300$~\cite{chasto15,chamut17}. For such fields, the equilibrium composition of the outer crust is significantly changed~\cite{chapav12,chamut17}.

At high enough pressure, neutrons drip out of nuclei, marking the transition to the inner crust, where neutron-proton clusters thus coexist with unbound neutrons. In the presence of a magnetic field, the onset of neutron emission is shifted to lower or higher densities depending on the magnetic field strength~\cite{chapav12,chasto15}. In the strongly quantizing regime, the neutron-drip pressure and density increase almost linearly with $B_\star$. Accurate analytical formulas can be found in Ref.~\cite{chasto15}. The role of Landau-Rabi quantization of electrons on the inner crust of a magnetar was previously studied in the Thomas-Fermi approximation~\cite{nandi2011}. The clusters were predicted to be larger and separated by a smaller distance than in the absence of magnetic fields, whereas the neutron liquid was found to be more dilute. 
However, these effects were  significant only for the extremely high value $B_\star=10^4$ of the magnetic field strength. 
Because nuclear shell effects were neglected in this study, the number of protons in clusters varied continuously with increasing density. However, shell effects are known to play a very important role in determining the equilibrium composition of unmagnetized neutron-star crusts, favoring layers made of clusters with specific `magic' proton numbers (see, e.g. Ref.~\cite{pearson2018}). 
Unlike the equation of state, elastic and transport properties are very sensitive to the crustal composition and this has implications for the seismic frequencies~\cite{nandi2016} and the cooling of the star~\cite{vigano2013,potchab2018}.

In this paper, we present new calculations of magnetar crusts based on the nuclear-energy density functional theory~\cite{duguet2014}.  
We have recalculated the properties of the outer crust using updated experimental atomic mass data supplemented with the Brussels-Montreal atomic mass model HFB-24~\cite{gor13}. Moreover, we have extended our investigations of highly-magnetized matter to the inner crust of a magnetar. 
To this end, we have employed the same functional BSk24 that underlies the HFB-24 model thus 
providing a unified and consistent treatment of both the outer and inner regions of the crust. Results are compared to those obtained for unmagnetized neutron stars using the same functional~\cite{pearson2018}. 

\section{Model of magnetar crusts}
\label{model}

\subsection{Outer crust}
We determine the properties of the outer crust of a magnetar adopting a model that we have previously presented in 
detail in Ref.~\cite{chapav12}. The crust is supposed to be stratified into different layers, each of which consists of a perfect 
crystal made of a single nuclear species embedded in a charge 
neutralizing relativistic electron Fermi gas that is fully degenerate. The lattice correction is taken into account but higher-order corrections, 
such as electron exchange and screening, are ignored. As in the absence of magnetic fields, nuclei are arranged on a body-centered 
cubic lattice~\cite{kozhberov2016}. However, the equilibrium composition can be altered due to the quantization of the electron motion 
perpendicular to the magnetic field. The energy levels of a relativistic electron 
Fermi gas in a magnetic field were first calculated by Rabi~\cite{rabi28}. Corresponding expressions for the electron pressure and 
energy density can be found, e.g., in Ref.~\cite{chapav12}. The equilibrium properties of any crustal layer at pressure $P$ and magnetic field 
strength $B_\star= B/B_{\rm rel}$ are determined by minimizing the Gibbs free energy per nucleon $g$ (see, e.g., Appendix of 
Ref.~\cite{chafant15}), given by \cite{chapav12}
\begin{equation}
g=\frac{M^\prime(A,Z)}{A}+\frac{Z}{A}\left( \mu_e - m_e c^2 +\frac{4}{3}C e^2 n_e^{1/3} Z^{2/3}\right)\, ,
\end{equation} 
\begin{equation}
P=P_e+\frac{1}{3} C e^2 n_e^{4/3}\, ,
\end{equation}
where $\mu_e$ is the electron Fermi energy, $P_e$ is the pressure of the ideal electron Fermi gas, $C\simeq -1.4442$ is the crystal structure constant~\cite{baiko01}, and $M^\prime(A,Z)$ is the nuclear mass (including the rest mass of nucleons and $Z$ electrons). The latter can 
be obtained from the atomic mass $M(A,Z)$ after subtracting out the binding 
energy of the atomic electrons (see Eq.~(A4) of Ref.~\cite{lpt03}). 
In principle, the presence of a high magnetic field may change the structure of nuclei~\cite{pen11,stein16}, inducing additional changes in the crustal 
composition~\cite{bas15}. However, complete nuclear mass tables accounting for the effects of the magnetic field are not yet available. We have therefore considered nuclear masses in the absence of magnetic fields. 

In our previous investigations~\cite{chapav12,chasto15,chamut17}, we made use of experimental atomic mass data from the 2012 Atomic Mass Evaluation (AME)~\cite{ame12}. In the present work, we have determined the equilibrium composition of the outer crust using the data from the 2016 AME~\cite{ame16} supplemented by recent mass measurements of copper isotopes~\cite{welker2017}, as in Ref.~\cite{pearson2018} for unmagnetized neutron stars. For the masses that have not been 
measured, we have adopted as in Ref.~\cite{pearson2018} the theoretical nuclear mass table HFB-24 from the BRUSLIB database\footnote{ http://www.astro.ulb.ac.be/bruslib/}. These masses
were obtained from self-consistent deformed Hartree-Fock-Bogoliubov (HFB) calculations using the generalized Skyrme functional BSk24~\cite{gor13}. This microscopic model was fitted to the 2353 measured masses of nuclei with neutron number $N \geq 8$ and proton number  $Z \geq 8 $ from the 2012 AME~\cite{ame12}, with a root-mean-square deviation of 0.549 MeV. This model provides an equally good fit to the 2408 measured masses of nuclei with $N$ and $Z \geq 8 $ from the 2016 AME. 

\subsection{Inner crust}

Ignoring neutron band-structure effects, the boundary between the outer 
and inner crust is determined by the condition $g=m_n c^2$, where $m_n$ is the neutron mass (see, e.g. Refs.~\cite{cfzh15,chasto15} for a detailed discussion). 
We have extended our studies of magnetar crusts to the inner part by implementing the effects of Landau-Rabi quantization in the computer code developed by the Brussels-Montreal collaboration~\cite{onsi2008,pearson2012,pearson2015}. This code is based on the fourth-order extended Thomas-Fermi method with
proton shell and pairing corrections added perturbatively using the Strutinsky integral theorem. This ETFSI method is 
a computationally very fast approximation to the HFB equations. Nuclear clusters are supposed to be unaffected by the presence of the magnetic field, and are 
further supposed to be spherical. This assumption may be questionable in the bottom layers of the crust, where nucleons could arrange themselves in so called nuclear ``pastas''. Indeed, instabilities in the densest part of unmagnetized neutron star crusts were previously reported within the same framework~\cite{pearson2018}. Since the magnetic fields that we consider here are not expected to have any significant effect on this crustal region, we restrict ourselves to  densities below $\bar{n}=0.07$ fm$^{-3}$. The Coulomb lattice is described following the approach of Wigner and Seitz. Nucleon density distributions in the Wigner-Seitz cell are parameterized as ($q = n, p$ for neutrons or protons respectively) 
\begin{equation}
n_q(r) = n_{B,q} + n_{\Lambda,q} f_q(r)\, ,
\end{equation}
in which $n_{B,q}$ are the background nucleon number densities, $n_{\Lambda,q}$ are nucleon number densities characterizing the clusters, and the cluster shape is described by the function 
\begin{equation}
f_q(r)=\Biggl\{1 + \exp\left[ \left( \frac{C_q-R_c}{r-R_c} \right)^2 -1 \right] \exp\left( \frac{r-C_q}{a_q}\right) \Biggr\}^{-1} \, ,
\end{equation}
where $C_q$ is the cluster radius defined as the half width at half maximum, $a_q$ the diffuseness of the cluster surface, and $R_c$ is the radius of the Wigner-Seitz cell. 
The numbers of protons and neutrons in the Wigner-Seitz cell are given by
\begin{equation}
Z= 4\pi  \int_0^{R_c} r^2 n_p(r){\rm d}r\, ,
\end{equation}
\begin{equation}
N= 4\pi \int_0^{R_c} r^2 n_n(r){\rm d}r\,,
\end{equation}
respectively. 

The equation of state of nuclear clusters and free neutrons is calculated 
from the same nuclear energy density functional BSk24~\cite{gor13}, as that underlying 
the nuclear mass model HFB-24 used in the outer crust. This functional was not only fitted to nuclear masses but was also 
constrained to reproduce the microscopic neutron-matter equation of state labelled `V18' in~\cite{ls08}, as obtained from 
the Brueckner Hartree-Fock approach using realistic two- and three-body forces. We thus believe that this functional is well 
suited for describing the neutron-rich matter of the inner crust of a neutron star. 
The analytical approximations implemented in the routines developed by Potekhin and Chabrier~\cite{eosmag} were adopted to calculate the equation of state of the 
cold magnetized electron Fermi gas. 

Minimizing the Gibbs free energy per nucleon at fixed pressure is numerically more delicate in the inner crust than in the outer crust because 
the pressure also depends on the density of free neutrons. Instead, it is more convenient to minimize the energy per nucleon $e$ at fixed  average 
baryon number density $\bar n$ (see, e.g., Ref.~\cite{pearson2012} for a discussion). We have thus followed the same approach as in our previous studies of unmagnetized neutron stars. The energy per nucleon is given by 
\begin{equation}
e=\frac{4\pi}{A}\int_0^{R_c}r^2 \mathcal{E}(r) {\rm d}r\, , 
\end{equation}
where $A=Z+N$ denotes the total number of nucleons in the Wigner-Seitz cell. The energy density $\mathcal{E}$ is calculated as in Ref.~\cite{pearson2018} except that the kinetic energy density of the electron Fermi gas is now determined taking into account the presence of a magnetic field. Similarly, the pressure is calculated as described in Ref.~\cite{pearson2018} but now including the magnetic field effects on the electron contribution.

\section{Equilibrium composition and equation of state of magnetar crusts}
\label{results}

\subsection{Outer crust}

The composition of the shallowest layers of the crust of a neutron star is completely determined by experimental atomic mass measurements. 
Detailed results for unmagnetized neutron stars can be found in Ref.~\cite{pearson2018}. The crust was found to be stratified in a succession of layers made of $^{56}$Fe, $^{62}$Ni, $^{64}$Ni, $^{66}$Ni, $^{86}$Kr, $^{84}$Se, $^{82}$Ge, and $^{80}$Zn with increasing pressure. In the deeper regions, the HFB-24 model predicts the occurrence of progressively more neutron rich nuclei that are less bound: $^{78}$Ni, $^{80}$Ni, $^{124}$Mo, $^{122}$Zr, $^{121}$Y, $^{120}$Sr, $^{122}$Sr, and $^{124}$Sr. At some pressure $P_{\rm drip}\simeq 4.87\times 10^{-4}$~MeV~fm$^{-3}$, neutrons drip out of the nuclei $^{124}$Sr, delimiting the boundary between the outer and inner regions of the crust. 

We have recalculated the equilibrium composition of the outer crust for different magnetic field strengths. Results are collected in Tables~\ref{tab:trans1000}, \ref{tab:trans2000} and \ref{tab:trans3000}. As in our earlier study~\cite{chapav12}, we find that the mean baryon number density at the stellar surface (where the pressure $P$ vanishes) is dramatically increased
due to magnetic condensation, reaching the value $\bar n_s \simeq 3.72\times 10^{-6}$ fm$^{-3}$ for $B_\star=3000$. The equation of state in the region beneath is well described by Eq.~(\ref{magnetic_cond}). 
When comparing our new results for the composition with the ones \cite{bjp2013,chamut17} obtained using the 2012 AME~\cite{ame12}, we notice several differences. First of all, the pressure between some adjacent crustal layers and their mean baryon densities are slightly shifted due to refined mass measurements. For instance, the mass excess of $^{56}$Fe now estimated as $-60.607$ MeV~\cite{ame16} instead of $-60.606$ MeV~\cite{ame12} leads to a small increase of the maximum pressure up to which $^{56}$Fe is present, from $P_{\rm max}=1.98\times 10^{-7}$ MeV~fm$^{-3}$ to $P_{\rm max}=2.00\times 10^{-7}$ MeV~fm$^{-3}$ for $B_\star=1000$. Likewise, the maximum density increases from $\bar{n}_{\rm max}=2.62\times 10^{-6}$ fm$^{-3}$ to $\bar{n}_{\rm max}=2.63\times 10^{-6}$ fm$^{-3}$. These changes can be understood from the approximate condition determining the transition between two adjacent layers made of nuclei ($A_1$, $Z_1$) and ($A_2$, $Z_2$) respectively~\cite{chamut17}
\begin{equation}\label{eq:threshold-condition}
\mu_e + C e^2 n_e^{1/3}\biggl(\frac{Z_2^{5/3}}{A_2} + \frac{1}{3}\frac{Z_2 Z_1^{2/3}}{A_2}-\frac{4}{3} \frac{Z_1^{5/3}}{A_1} \biggr)
\left(\frac{Z_2}{A_2}-\frac{Z_1}{A_1}\right)^{-1} = \mu_e^{1\rightarrow2}\, ,
\end{equation}
where
\begin{equation}\label{eq:mue12}
\mu_e^{1\rightarrow2}\equiv \left(\frac{M^\prime(A_1,Z_1)c^2}{A_1}-\frac{M^\prime(A_2,Z_2)c^2}{A_2}\right)\left(\frac{Z_2}{A_2}-\frac{Z_1}{A_1}\right)^{-1} +  m_e c^2\, .
\end{equation}
The layer lying beneath that made of $^{56}$Fe contains $^{62}$Ni. The slight decrease of the mass of $^{56}$Fe thus entails an increase of $\mu_e^{1\rightarrow2}$, hence also of $n_e$ and of $P$. Revisions in the atomic mass estimates are more significant for exotic nuclei. In particular, 
the mass excess of $^{130}$Cd, which was previously estimated as $-61.534$ MeV~\cite{ame12} has been revised to $-61.118$ MeV~\cite{ame16}. Making use of the 2012 AME with the HFB-24 mass model, this nuclide was predicted to appear in the crust of a neutron star for $B_\star=1697$~\cite{chamut17}. With the 2016 AME, $^{130}$Cd is no longer favored up to the highest field considered. Since the publication of our previous works, new masses of neutron-rich nuclei have been measured. These include that of $^{79}$Cu~\cite{welker2017}, which was previously estimated using the mass model HFB-24. This nuclide, which was found up to $B_\star=1617$~\cite{chamut17}, is now disfavored. For the highest magnetic fields considered here namely $B_\star=3000$, we observe that $^{121}$Y and $^{80}$Zn have disappeared, whereas $^{124}$Zr has appeared.

\begin{table}
\centering
\caption{Composition of the outer crust of a cold nonaccreted neutron
star endowed with a magnetic field $ B_{\star} = 1000$. Results were
obtained using experimental masses from the 2016 Atomic Mass Evaluation
\cite{ame16} supplemented with the Brussels-Montreal nuclear mass model
HFB-24 \cite{gor13}. The mean baryon number densities $\bar n$ are
measured in units of fm$^{-3}$ and the transition pressures $ P_{1 \to 2}
$ are in units of MeV fm$^{-3}$. Properties in the upper part of the
table are fully determined by experimental atomic masses. See text for
details.} 
\label{tab:trans1000} 
\begin{tabular}{clcllll} 
\hline
$Z$ & $A$ & $\bar{n}_{\rm min}$  & $ \bar{n}_{\rm max}$ & $P_{1\to2}$  \\ 
\hline
26 & 56  & 9.96$\times$ $10^{-7}$ & 2.63 $\times$ $10^{-6}$ & 2.00 $\times$ $10^{-7}$   \\ 
28 & 62  & 2.72 $\times$ $10^{-6}$ & 1.10 $\times$ $10^{-5}$ & 6.23 $\times$ $10^{-6}$   \\ 
28 & 64  & 1.14 $\times$ $10^{-5}$ & 1.42 $\times$ $10^{-5}$ & 1.01 $\times$ $10^{-5}$   \\
38 & 88  & 1.45 $\times$ $10^{-5}$ & 1.55 $\times$ $10^{-5}$ & 1.16 $\times$ $10^{-5}$   \\
36 & 86  & 1.60 $\times$ $10^{-5}$ & 2.60 $\times$ $10^{-5}$ & 3.21 $\times$ $10^{-5}$   \\
34 & 84  & 2.69 $\times$ $10^{-5}$ & 3.88 $\times$ $10^{-5}$ & 6.82 $\times$ $10^{-5}$   \\
32 & 82  & 4.02 $\times$ $10^{-5}$ & 5.22 $\times$ $10^{-5}$ & 1.16 $\times$ $10^{-4}$   \\
30 & 80  & 5.43 $\times$ $10^{-5}$ & 6.71 $\times$ $10^{-5}$ & 1.78 $\times$ $10^{-4}$   \\
\hline 
28 & 78  & 7.01 $\times$ $10^{-5}$ & 8.46 $\times$ $10^{-5}$ & 2.61 $\times$ $10^{-4}$   \\
28 & 80  & 8.68 $\times$ $10^{-5}$ & 8.96 $\times$ $10^{-5}$ & 2.79 $\times$ $10^{-4}$   \\
42 & 124 & 9.30 $\times$ $10^{-5}$ & 1.07 $\times$ $10^{-4}$ & 3.68 $\times$ $10^{-4}$   \\
40 & 122 & 1.10 $\times$ $10^{-4}$ & 1.18 $\times$ $10^{-4}$ & 4.25 $\times$ $10^{-4}$   \\
39 & 121 & 1.20 $\times$ $10^{-4}$ & 1.21 $\times$ $10^{-4}$ & 4.32 $\times$ $10^{-4}$   \\
38 & 120 & 1.23 $\times$ $10^{-4}$ & 2.18 $\times$ $10^{-4}$ & 5.06 $\times$ $10^{-4}$   \\
38 & 122 & 2.22 $\times$ $10^{-4}$ & 2.74 $\times$ $10^{-4}$ & 6.20 $\times$ $10^{-4}$   \\
38 & 124 & 2.79 $\times$ $10^{-4}$ & 2.91 $\times$ $10^{-4}$ & 6.56 $\times$ $10^{-4}$   \\
\end{tabular} 
\end{table}

\begin{table}
\centering
\caption{Same as Table~\ref{tab:trans1000} for $B_\star=2000$.}
\label{tab:trans2000}
\begin{tabular}{clcllll}
\hline
$Z$ & $A$ & $\bar{n}_{\rm min}$ & $ \bar{n}_{\rm max}$  & $P_{1\to2}$ \\ 
\hline
26 & 56  & 2.29 $\times$ $10^{-6}$ & 5.05 $\times$ $10^{-6}$ & 3.15 $\times 10^{-7}$    \\
28 & 62  & 5.24 $\times$ $10^{-6}$ & 2.22 $\times$ $10^{-5}$ & 1.23 $\times$ $10^{-5}$    \\
38 & 88  & 2.34 $\times$ $10^{-5}$ & 3.37 $\times$ $10^{-5}$ & 2.68 $\times$ $10^{-5}$     \\
36 & 86  & 3.47 $\times$ $10^{-5}$ & 5.49 $\times$ $10^{-5}$ & 7.06 $\times$ $10^{-5}$     \\
34 & 84  & 5.67 $\times$ $10^{-5}$ & 8.08 $\times$ $10^{-5}$ & 1.46 $\times$ $10^{-4}$     \\
32 & 82  & 8.37 $\times$ $10^{-5}$ & 1.08 $\times$ $10^{-4}$ & 2.44 $\times$ $10^{-4}$   \\
50 & 132 & 1.12 $\times$ $10^{-4}$ & 1.15 $\times$ $10^{-4}$ & 2.58 $\times$ $10^{-4}$    \\
30 & 80  & 1.15 $\times$ $10^{-4}$ & 1.32 $\times$ $10^{-4}$ & 3.40 $\times$ $10^{-4}$   \\
\hline
46 & 128 & 1.38 $\times$ $10^{-4}$ & 1.71 $\times$ $10^{-4}$ & 5.22 $\times$ $10^{-4}$    \\
44 & 126 & 1.75 $\times$ $10^{-4}$ & 1.85 $\times$ $10^{-4}$ & 5.82 $\times$ $10^{-4}$    \\
42 & 124 & 1.91 $\times$ $10^{-4}$ & 2.19 $\times$ $10^{-4}$ & 7.69 $\times$ $10^{-4}$    \\
40 & 122 & 2.26 $\times$ $10^{-4}$ & 2.42 $\times$ $10^{-4}$ & 8.81 $\times$ $10^{-4}$   \\
40 & 124 & 2.46 $\times$ $10^{-4}$ & 2.48 $\times$ $10^{-4}$ & 8.99 $\times$ $10^{-4}$   \\
38 & 120 & 2.52 $\times$ $10^{-4}$ & 2.64 $\times$ $10^{-4}$ & 9.85 $\times$ $10^{-4}$   \\
38 & 122 & 2.69 $\times$ $10^{-4}$ & 2.85 $\times$ $10^{-4}$ & 1.11 $\times$ $10^{-3}$   \\
38 & 124 & 2.90 $\times$ $10^{-4}$ & 2.95 $\times$ $10^{-4}$ & 1.15 $\times$ $10^{-3}$  \\
\end{tabular}
\end{table}

\begin{table}
\centering
\caption{Same as Table~\ref{tab:trans1000} for $B_\star=3000$.} 
\label{tab:trans3000} 
\begin{tabular}{clcllll} 
\hline
$Z$ & $A$ & $\bar{n}_{\rm min}$ & $ \bar{n}_{\rm max}$ & $P_{1\to2}$ \\ 
\hline
26 & 56  & 3.72 $\times$ $10^{-6}$ & 7.36 $\times$ $10^{-6}$ & 3.89 $\times$ $10^{-7}$   \\ 
28 & 62  & 7.65 $\times$ $10^{-6}$ & 3.11 $\times$ $10^{-5}$ & 1.57 $\times$ $10^{-5}$   \\ 
38 & 88  & 3.30 $\times$ $10^{-5}$ & 5.33 $\times$ $10^{-5}$ & 4.45 $\times$ $10^{-5}$   \\
36 & 86  & 5.48 $\times$ $10^{-5}$ & 8.55 $\times$ $10^{-5}$ & 1.13 $\times$ $10^{-4}$   \\
34 & 84  & 8.83 $\times$ $10^{-5}$ & 1.24 $\times$ $10^{-4}$ & 2.30 $\times$ $10^{-4}$   \\
32 & 82  & 1.29 $\times$ $10^{-4}$ & 1.50 $\times$ $10^{-4}$ & 3.13 $\times$ $10^{-4}$   \\
50 & 132 & 1.56 $\times$ $10^{-4}$ & 2.01 $\times$ $10^{-4}$ & 5.26 $\times$ $10^{-4}$   \\
\hline 
46 & 128 & 2.11 $\times$ $10^{-4}$ & 2.61 $\times$ $10^{-4}$ & 8.13 $\times$ $10^{-4}$   \\
44 & 126 & 2.69 $\times$ $10^{-4}$ & 2.83 $\times$ $10^{-4}$ & 9.03 $\times$ $10^{-4}$   \\
42 & 124 & 2.91 $\times$ $10^{-4}$ & 3.34 $\times$ $10^{-4}$ & 1.19 $\times$ $10^{-3}$   \\
40 & 122 & 3.44 $\times$ $10^{-4}$ & 3.66 $\times$ $10^{-4}$ & 1.34 $\times$ $10^{-3}$   \\
40 & 124 & 3.72 $\times$ $10^{-4}$ & 3.80 $\times$ $10^{-4}$ & 1.40 $\times$ $10^{-3}$   \\
38 & 120 & 3.87 $\times$ $10^{-4}$ & 3.99 $\times$ $10^{-4}$ & 1.50 $\times$ $10^{-3}$   \\
38 & 122 & 4.06 $\times$ $10^{-4}$ & 4.31 $\times$ $10^{-4}$ & 1.69 $\times$ $10^{-3}$   \\
38 & 124 & 4.38 $\times$ $10^{-4}$ & 4.45 $\times$ $10^{-4}$ & 1.74 $\times$ $10^{-3}$   \\
\end{tabular} 
\end{table}

\subsection{Neutron-drip transition}

As discussed in our previous studies (see, e.g. Ref.~\cite{chasto15}), the onset of neutron drip is shifted to much higher pressures
in the presence of a strongly quantizing magnetic field. In particular, we have found that the neutron-drip transition occurs at 
pressure $1.74\times 10^{-3}$ MeV~fm$^{-3}$ for $B_\star=3000$ instead of $4.87\times 10^{-4}$~MeV~fm$^{-3}$ in the absence of magnetic field.
In comparison, the neutron-drip density is only moderately increased from $2.56\times10^{-4}$ fm$^{-3}$ to $4.45\times 10^{-4}$ fm$^{-3}$. 
As found in our previous works, the equilibrium nuclide at the neutron-drip point remains $^{124}$Sr independently of the magnetic field strength.

\subsection{Inner crust}

Determining the equilibrium properties of the inner crust can be numerically tricky. The energy minimum may be not only very flat in parameter space, but differences between local minima also tend to decrease with increasing density (see e.g. Ref.~\cite{pearson2018}). The occurrence of quantum oscillations, especially in the weakly quantizing regime in which many Landau-Rabi levels are filled (see, e.g., Ref.~\cite{chamut17}), adds to the difficulty. 

\begin{figure}[ht]
\begin{center}
\includegraphics[width=0.75\textwidth]{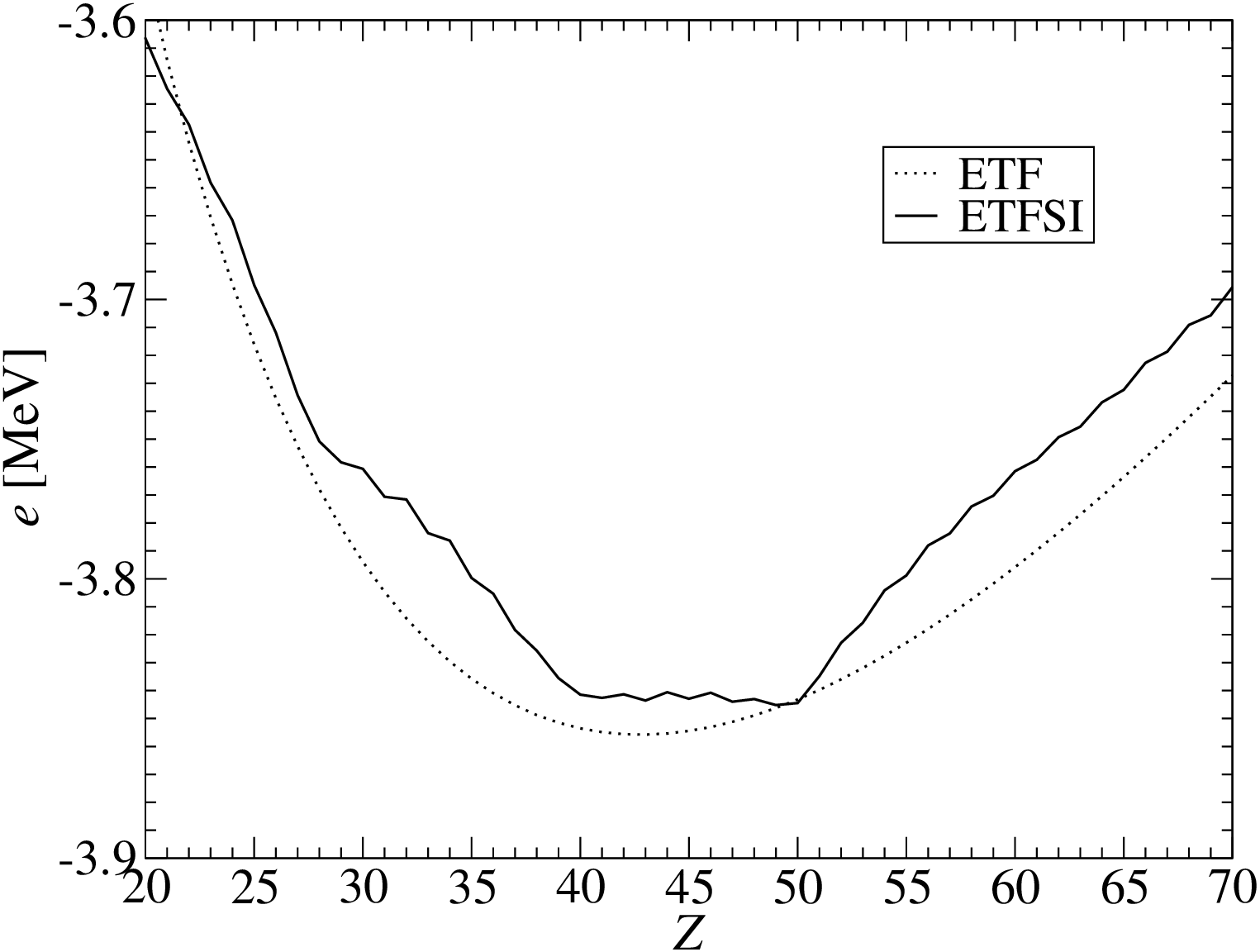}
\caption{Energy per nucleon $e$ with the neutron rest mass energy subtracted (in MeV) as a function of the proton number $Z$ in the inner crust of a magnetar for $\bar{n}=4.45\times10^{-4}$ fm$^{-3}$ and $B_\star=3000$. Fourth-order extended Thomas-Fermi calculations with and without proton shell and pairing corrections are shown by the solid and dotted lines respectively. See text for details.}
\label{fig:e_vs_Z}
\end{center}
\end{figure}

To test our inner crust code, we have computed the properties of different crustal layers in the limit of a relatively low magnetic field by setting $B_\star=1$. As shown in Table~\ref{tab:b0vsb1}, the results for weakly magnetized neutron stars are in very good agreement with those obtained with the code for strictly unmagnetized neutron stars. The small deviations could be attributed to the interpolations implemented in the routines of Ref.~\cite{eosmag}. To further test our inner crust code, we have compared the neutron-drip properties with those obtained using our outer crust code. Results are summarized in Table~\ref{tab:ndrip_comp}. Whereas the discrepancies for the pressure and the energy per nucleon are found to be of the same order as in the absence of magnetic fields, those for the proton number $Z$ and neutron number $N$ appear to be larger especially for $B_\star=3000$. As discussed in Ref.~\cite{pearson2012}, we do not expect a perfect matching between the two codes since the outer and inner regions of the crust are described using different approximations. Moreover, we cannot exclude the possibility that our inner crust code converged to a local minimum energy configuration instead of the true equilibrium state. Indeed, the numerical search for the absolute minimum is made difficult by the extremely small energy differences that can exist between different configurations. In particular, the energy per nucleon at the neutron-drip point for $B_\star=3000$ appears to be very flat for $Z$ lying between 40 and 50, as illustrated in Fig.~\ref{fig:e_vs_Z}. The configuration with $Z=49$ differs by about 2.5 keV per nucleon from the configuration with $Z=41$. The figure also illustrates the importance of proton shell and pairing corrections for determining the equilibrium composition. 

\begin{table}
\caption{Inner crust properties at different mean baryon number densities $\bar n$ in fm$^{-3}$, as obtained with the two versions of 
our inner crust code: the one for magnetars with $B_\star=1$ and the other for unmagnetized neutron stars (in parentheses). 
In all cases, the proton number in the Wigner-Seitz cell 
was found to be $Z=40$. $P$ is the pressure in MeV~fm$^{-3}$ and $e$ is the energy per nucleon in MeV (with the neutron rest mass energy subtracted). See text for details.}
\label{tab:b0vsb1}
\centering
\begin{tabular}{ccc}
\hline  
$\bar n$  & $P$ & $e$  \\
\hline  
3.03750$\times 10^{-4}$ &  5.46740$\times 10^{-4}$ (5.47181$\times 10^{-4}$) & -1.52400$\times 10^{0}$ (-1.52387$\times 10^{0}$) \\
5.47367$\times 10^{-4}$ &  7.47651$\times 10^{-4}$ (7.49487$\times 10^{-4}$) & -6.54916$\times 10^{-1}$ (-6.54893$\times 10^{-1}$) \\
9.86374$\times 10^{-4}$ &  1.17275$\times 10^{-3}$ (1.17361$\times 10^{-3}$) & 4.98938$\times 10^{-2}$ (4.99243$\times 10^{-2}$) \\
1.77748$\times 10^{-3}$ &  2.15441$\times 10^{-3}$ (2.15216$\times 10^{-3}$) & 7.20421$\times 10^{-1}$ (7.20375$\times 10^{-1}$) \\
3.20307$\times 10^{-3}$ &  4.43514$\times 10^{-3}$ (4.42966$\times 10^{-3}$) & 1.45537$\times 10^{0}$ (1.45530$\times 10^{0}$) \\
5.77204$\times 10^{-3}$ &  9.60512$\times 10^{-3}$ (9.60349$\times 10^{-3}$) & 2.32853$\times 10^{0}$ (2.32851$\times 10^{0}$) \\
1.04014$\times 10^{-2}$ &  2.08081$\times 10^{-2}$ (2.07968$\times 10^{-2}$) & 3.38692$\times 10^{0}$ (3.38697$\times 10^{0}$) \\
1.87437$\times 10^{-2}$ &  4.34014$\times 10^{-2}$ (4.33952$\times 10^{-2}$) & 4.63934$\times 10^{0}$ (4.63936$\times 10^{0}$) \\
3.37767$\times 10^{-2}$ & 8.60910$\times 10^{-2}$ (8.60561$\times 10^{-2}$) & 6.04953$\times 10^{0}$ (6.04951$\times 10^{0}$) \\
6.08667$\times 10^{-2}$ &  1.77762$\times 10^{-1}$ (1.77714$\times 10^{-1}$) & 7.60048$\times 10^{0}$ (7.60046$\times 10^{0}$) \\
\hline  
\end{tabular}
\end{table}

\begin{table}
\caption{Comparison between our outer- and inner-crust codes at the neutron-drip point; results for the former are indicated in parentheses. $B_\star$ is the magnetic field strength,  $\bar n_{\rm drip}$ is the mean baryon number density at the neutron drip-point in fm$^{-3}$, $e_{\rm drip}$ is the corresponding energy per nucleon in MeV (with the neutron rest mass energy subtracted), $P_{\rm drip}$ is the corresponding pressure in MeV~fm$^{-3}$, $Z$ and $N$ are the corresponding proton and neutron numbers respectively.  
}  
\label{tab:ndrip_comp}
\centering
\begin{tabular}{rccccc}
\hline  
$B_\star$ & $\bar n_{\rm drip}$  & $Z$ & $N$ & $e_{\rm drip}$  & $P_{\rm drip}$  \\
\hline  
1000  & 2.91$\times 10^{-4}$ & 41 (38) & 96 (86) & -2.17 (-2.25) & 6.65$\times 10^{-4}$ (6.56$\times 10^{-4}$) \\
2000  & 2.95$\times 10^{-4}$ & 41 (38) & 95 (86) & -3.82 (-3.90) &1.14$\times 10^{-3}$ (1.15$\times 10^{-3}$) \\
3000  & 4.45$\times 10^{-4}$ & 49 (38) & 113 (86) & -3.85 (-3.92) &1.75$\times 10^{-3}$ (1.74$\times 10^{-3}$) \\
\hline  
\end{tabular}
\end{table}

\begin{figure}[ht]
\begin{center} 
\includegraphics[width=0.75\textwidth]{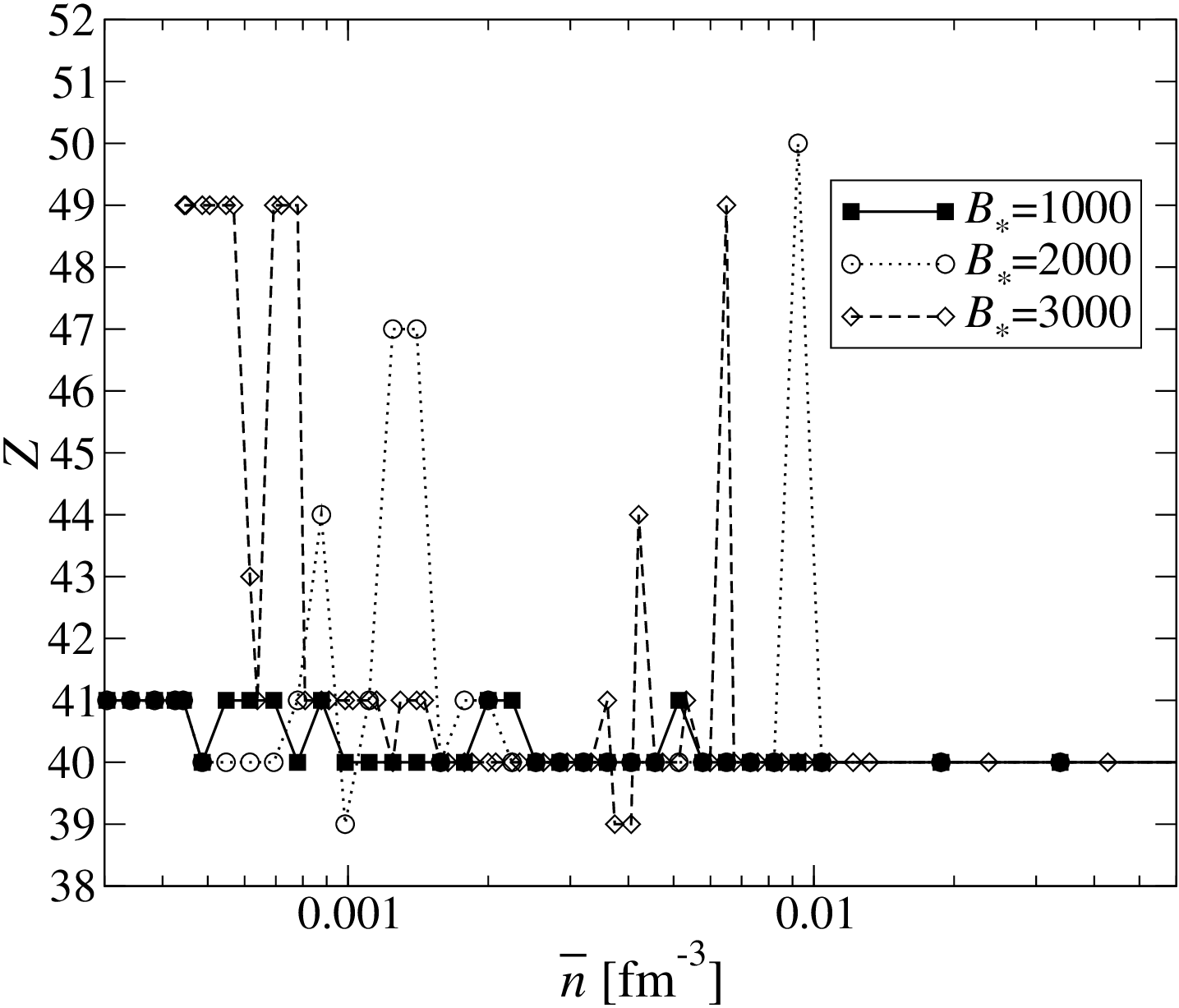}
\caption{Proton number $Z$ in the Wigner-Seitz cell as a function of mean baryon density $\bar{n}$ in fm$^{-3}$ in the inner crust of a magnetar for different magnetic field strengths. In the absence of magnetic field, $Z=40$ for all densities. 
}
\label{fig:Z_vs_n}
\end{center}
\end{figure}

\begin{figure}[ht]
\begin{center} 
\includegraphics[width=0.75\textwidth]{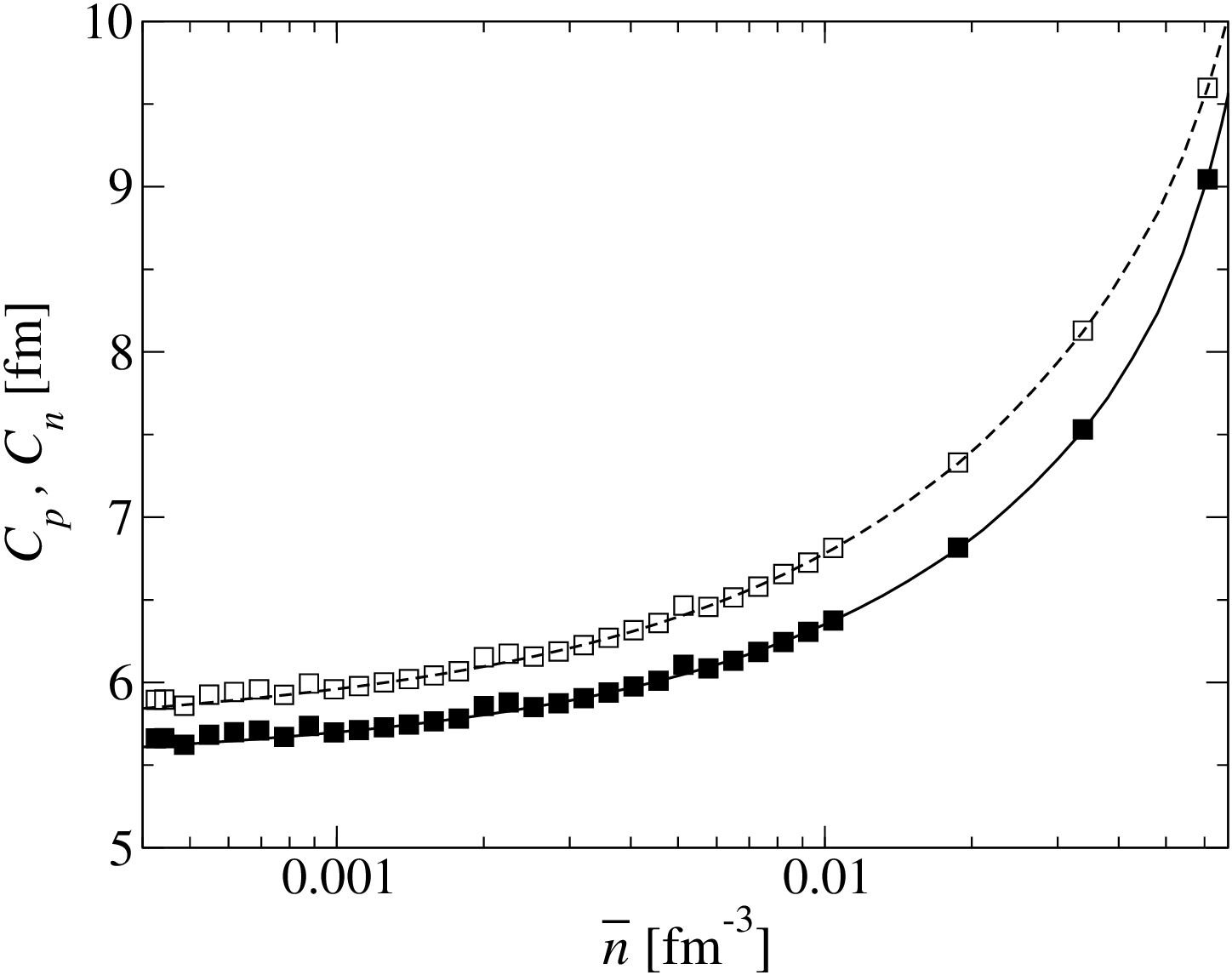}
\caption{Cluster size in fm as a function of mean baryon density $\bar{n}$ in fm$^{-3}$ in the inner crust of a magnetar for $B_\star=1000$ (square symbols) and for $B_\star=0$ (lines). The cluster radius $C_n$ as determined by neutron distribution is shown by empty symbols and dashed line, while the cluster radius $C_p$ as determined by the proton distribution is shown by filled symbols and solid line. 
}
\label{fig:CpCn_bs1000}
\end{center}
\end{figure}

\begin{figure}[ht]
\begin{center} 
\includegraphics[width=0.75\textwidth]{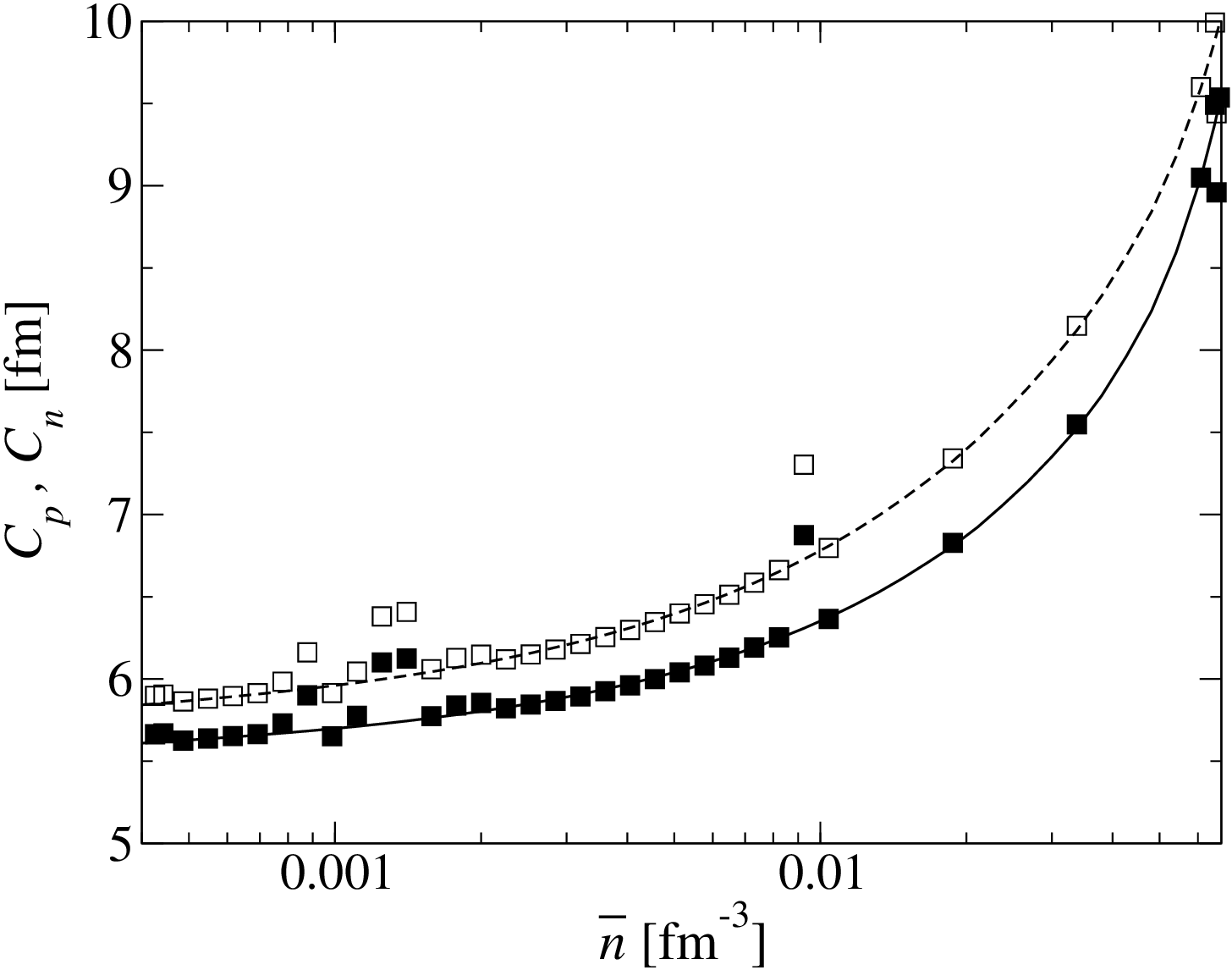}
\caption{Same as Fig.~\ref{fig:CpCn_bs1000} for $B_\star=2000$ (square symbols) and for $B_\star=0$ (lines). 
}
\label{fig:CpCn_bs2000}
\end{center}
\end{figure}

\begin{figure}[ht]
\begin{center} 
\includegraphics[width=0.75\textwidth]{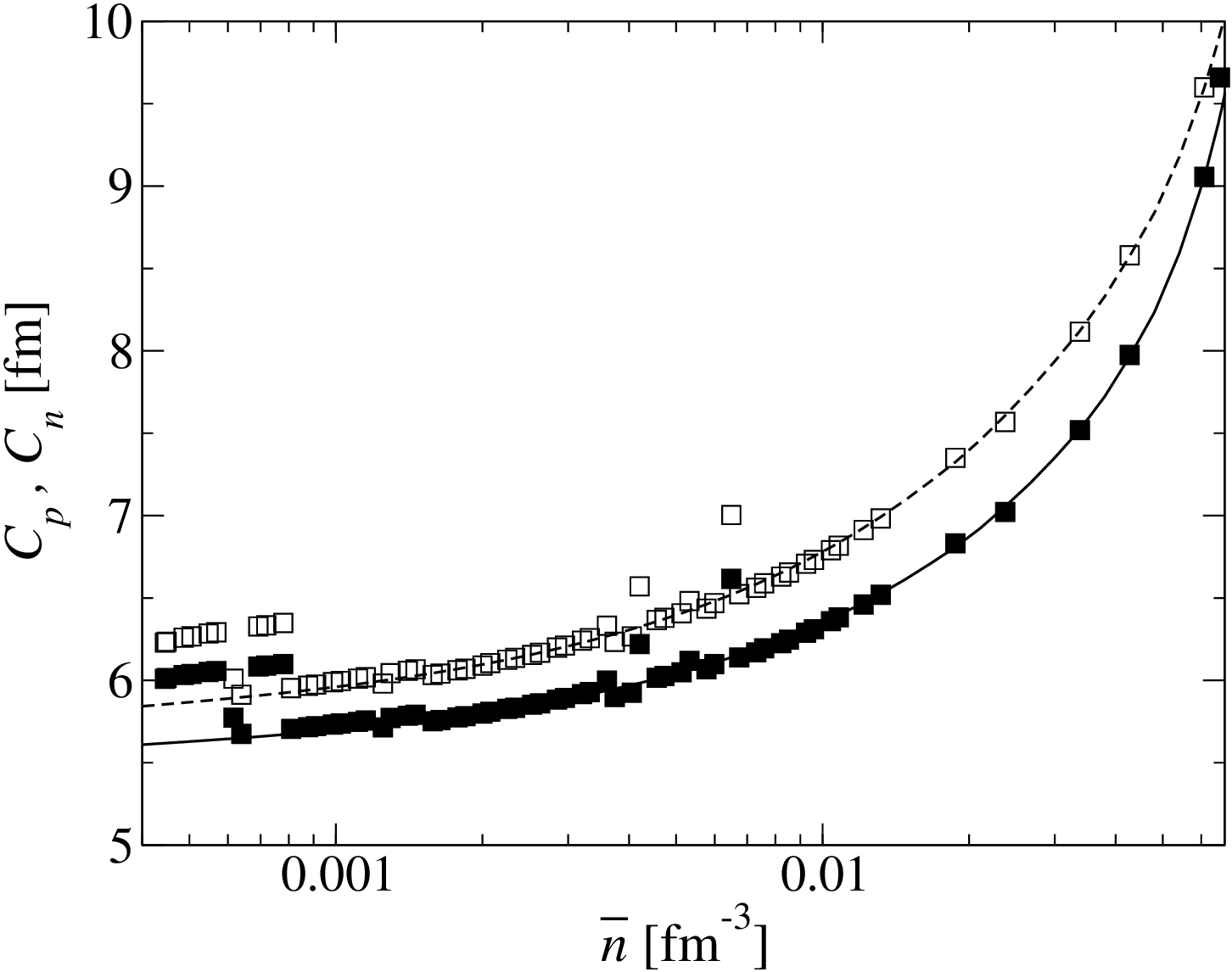}
\caption{Same as Fig.~\ref{fig:CpCn_bs1000} for $B_\star=2000$ (square symbols) and for $B_\star=0$ (lines). 
}
\label{fig:CpCn_bs3000}
\end{center}
\end{figure}

\begin{figure}[ht]
\begin{center} 
\includegraphics[width=0.75\textwidth]{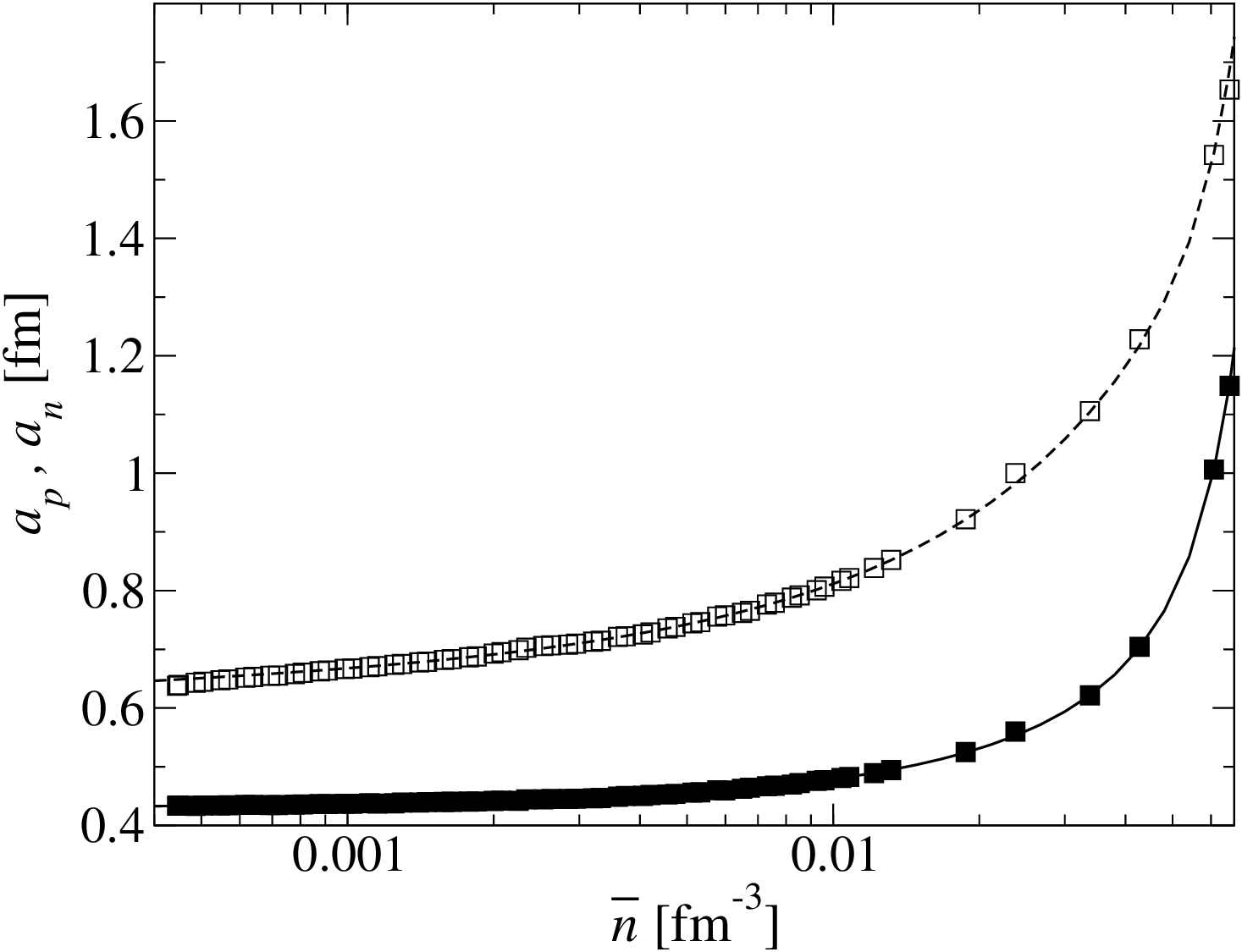}
\caption{Cluster surface diffuseness in fm as a function of mean baryon density $\bar{n}$ in fm$^{-3}$ in the inner crust of a magnetar for $B_\star=3000$ (square symbols) and for $B_\star=0$ (lines). The neutron diffuseness $a_n$ is shown by empty symbols and dashed line, while the proton diffuseness $a_p$ is shown by filled symbols and solid line. 
}
\label{fig:apan_bs3000}
\end{center}
\end{figure}

\begin{figure}[ht]
\begin{center} 
\includegraphics[width=0.75\textwidth]{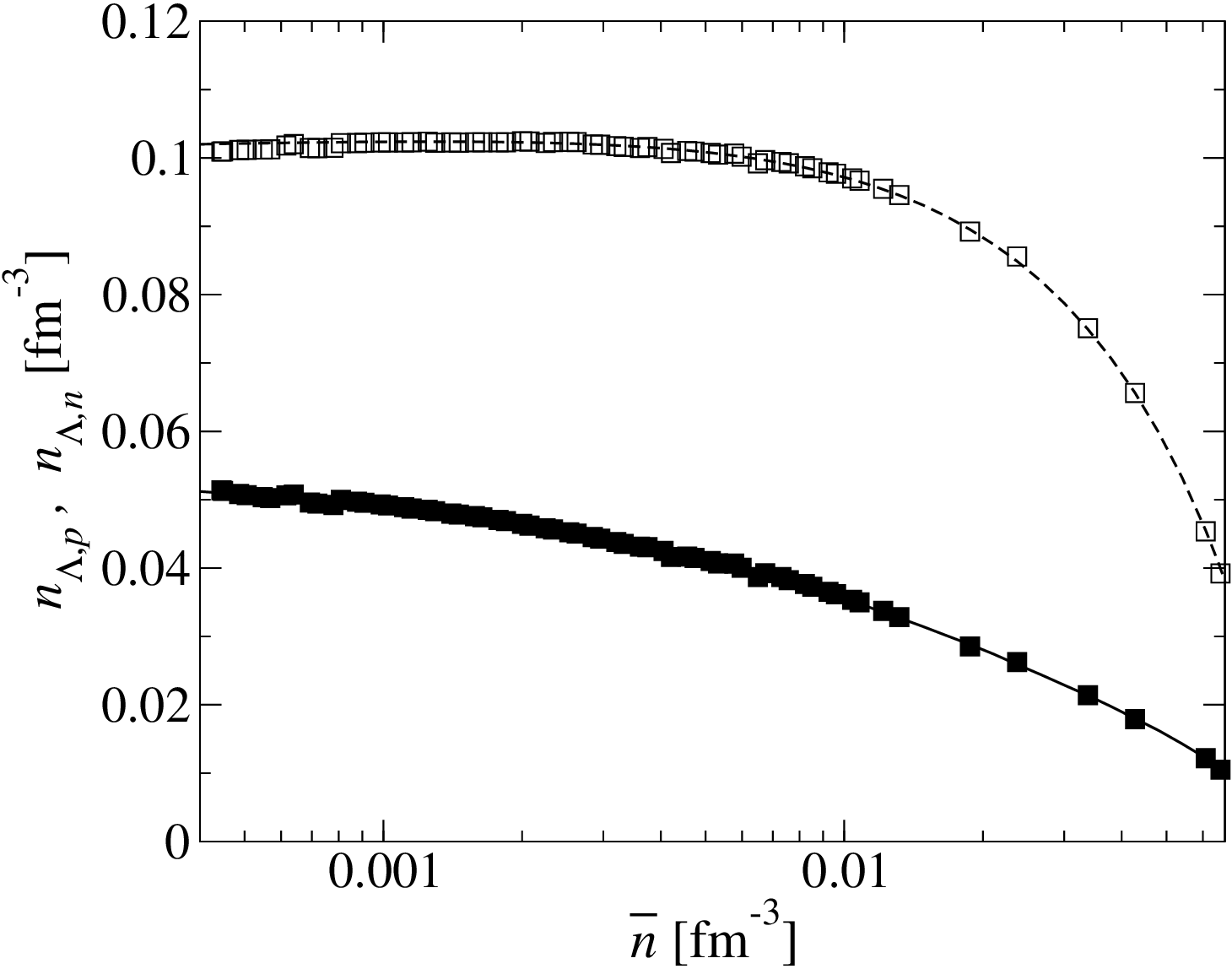}
\caption{Cluster densities in fm$^{-3}$ as a function of mean baryon density $\bar{n}$ in fm$^{-3}$ in the inner crust of a magnetar for $B_\star=3000$ (square symbols) and for $B_\star=0$ (solid line). The neutron density $n_{\Lambda,n}$ is shown by empty symbols and dashed line, while the proton density $n_{\Lambda,p}$ is shown by filled symbols and solid line. 
}
\label{fig:nLpnLn_bs3000}
\end{center}
\end{figure}

\begin{figure}[ht]
\begin{center} 
\includegraphics[width=0.75\textwidth]{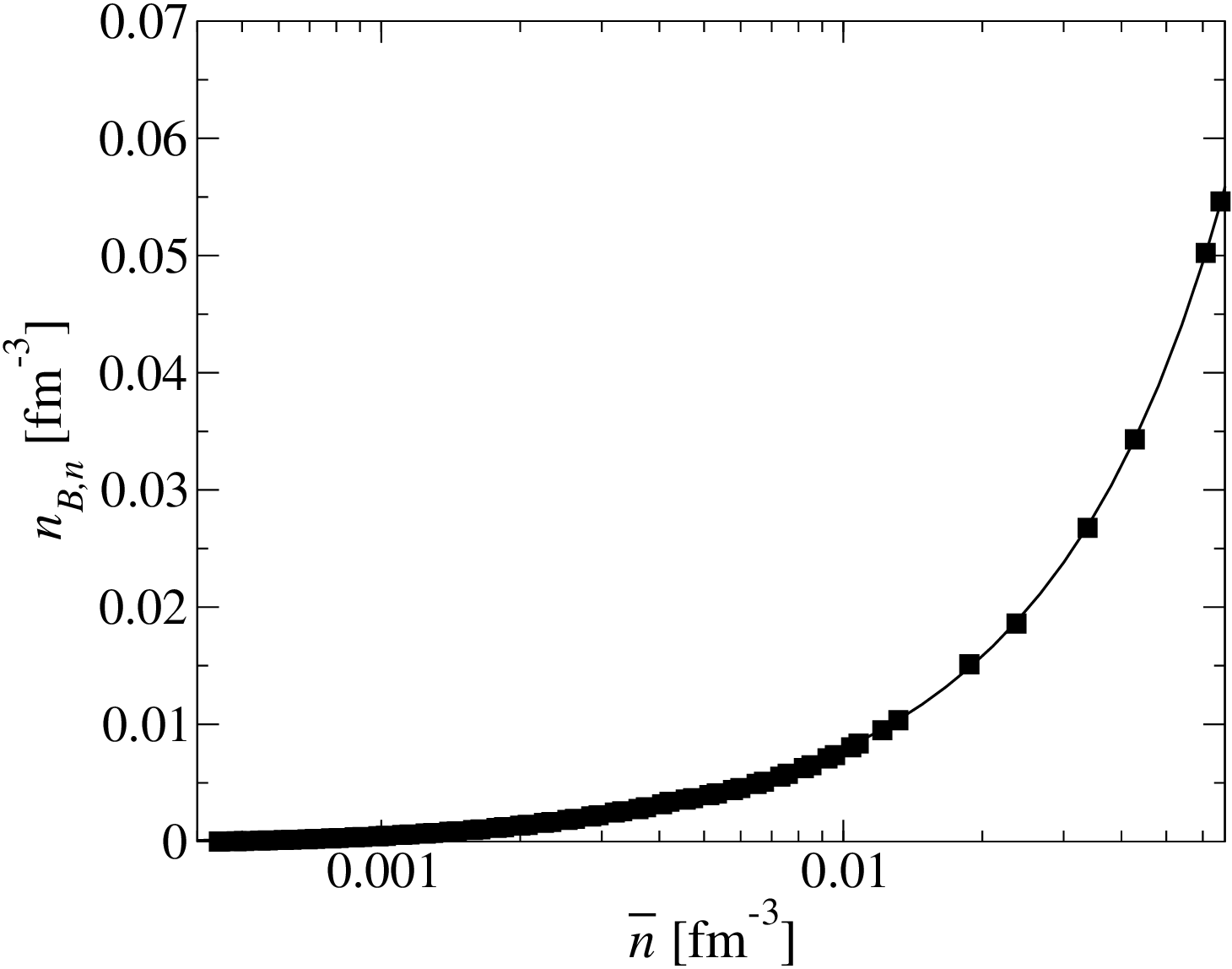}
\caption{Density $n_{B,n}$ of the neutron in fm$^{-3}$ as a function of mean baryon density $\bar{n}$ in fm$^{-3}$ in the inner crust of a magnetar for $B_\star=3000$ (square symbols) and for $B_\star=0$ (solid line).
}
\label{fig:nBn_bs3000}
\end{center}
\end{figure}

In the absence of magnetic fields, most regions of the inner crust of a neutron star were previously found to be made of nuclear clusters with $Z=40$ up to a mean density $\bar{n}=0.067$ fm$^{-3}$~\cite{pearson2018}. The inner crust of a magnetar is still  predicted to contain nuclear clusters with $Z=40$ but other values of $Z$ also appear to be energetically favored in some layers depending on the magnetic field strength, as shown in Fig.~\ref{fig:Z_vs_n}. Most remarkably, high-enough magnetic fields lead to the appearance of odd values of $Z$ around the `magic' values 40 and 50. The discontinuous changes in $Z$ with density $\bar{n}$ are reflected in the cluster size, as shown in Figs.~\ref{fig:CpCn_bs1000}, \ref{fig:CpCn_bs2000} and \ref{fig:CpCn_bs3000}. However, the cluster shape, as characterized by the diffusenesses $a_q$ and the densities $n_{\Lambda,q}$, remain unaltered, as can be seen in Figs.~\ref{fig:apan_bs3000} and \ref{fig:nLpnLn_bs3000}.  Likewise, the magnetic field is not found to have any significant effect on the density $n_{B,n}$ of the neutron liquid, as show in Fig.~\ref{fig:nBn_bs3000}. Since these conclusions hold for all three considered magnetic field strengths, only the results for $B_\star=3000$ have been plotted. 
 
\begin{figure}[ht]
\begin{center} 
\includegraphics[width=0.75\textwidth]{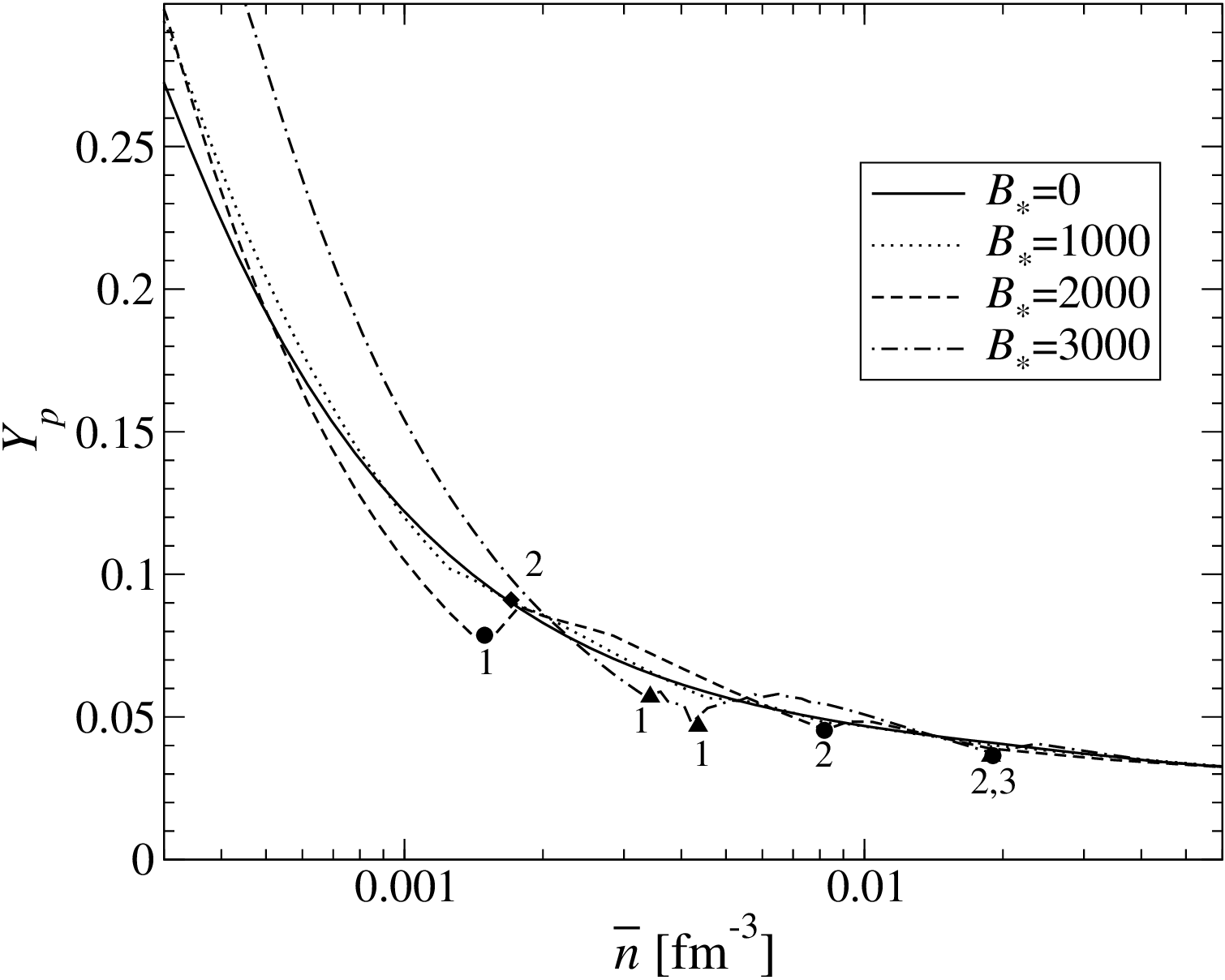}
\caption{Proton fraction $Y_p$ as a function of mean baryon density $\bar{n}$ in fm$^{-3}$ in the inner crust of a magnetar for different magnetic field strength $B_\star$.
}
\label{fig:Y_vs_n-inner}
\end{center}
\end{figure}

The mean proton fraction $Y_p=Z/A$, which is plotted in Fig.~\ref{fig:Y_vs_n-inner}, varies smoothly with $\bar{n}$ and 
exhibits typical fluctuations due to Landau-Rabi quantization. It should be remarked that the filling of Landau-Rabi levels does not necessarily increase monotonically with pressure. This stems from the fact that the composition also changes. For instance, in descending towards deeper layers in the crust for $B_\star=3000$, the lowest Landau-Rabi level is first fully populated at density $\bar{n}\simeq 0.0036$ fm$^{-3}$ corresponding to the electron Fermi energy $\mu_e=m_e c^2\sqrt{1+2 B_\star}\simeq 39.6$ MeV. However, further compression leads to a slight decrease of $\mu_e$ below the threshold. The lowest level is fully occupied again at density $\bar{n}\simeq 0.0046$ fm$^{-3}$. This leads to additional variations of $Y_p$ in this density range, as can be seen in Fig.~\ref{fig:Y_vs_n-inner}. At high enough densities, $Y_p$ becomes essentially independent of the magnetic field strength and matches that obtained for unmagnetized neutron stars.

\begin{figure}[ht]
\begin{center}
\includegraphics[width=0.75\textwidth]{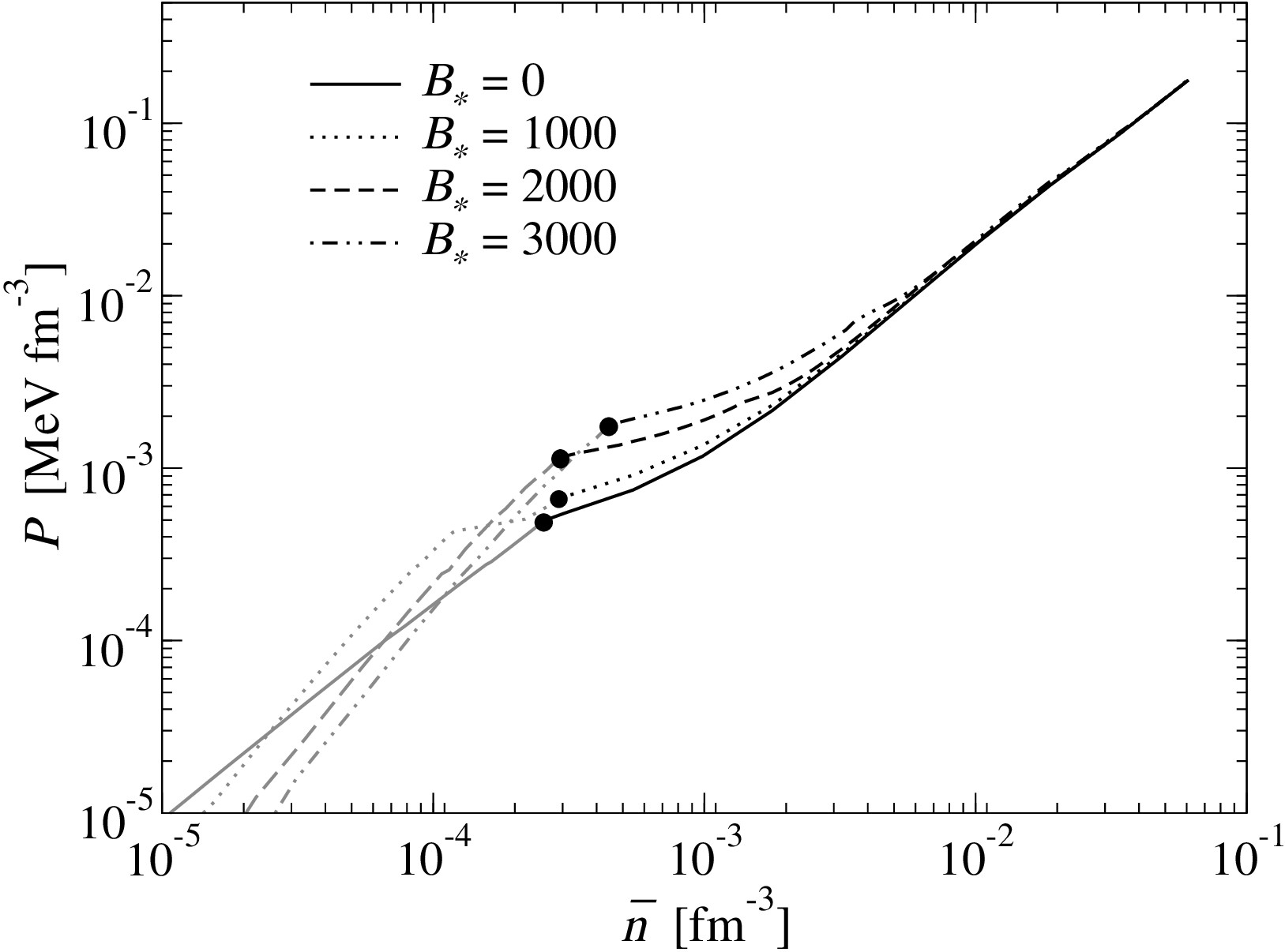}
\caption{Pressure $P$ in MeV~fm$^{-3}$ as a function of the mean baryon number density $\bar n$ in fm$^{-3}$ in the crust of a magnetar for different magnetic field strengths $B_\star$. The black circles mark the neutron-drip points. Results in grey are for the outer crust while those in black are for the inner crust.}
\label{fig:P_vs_n}
\end{center}
\end{figure}

\begin{figure}[ht]
\begin{center}
\includegraphics[width=0.75\textwidth]{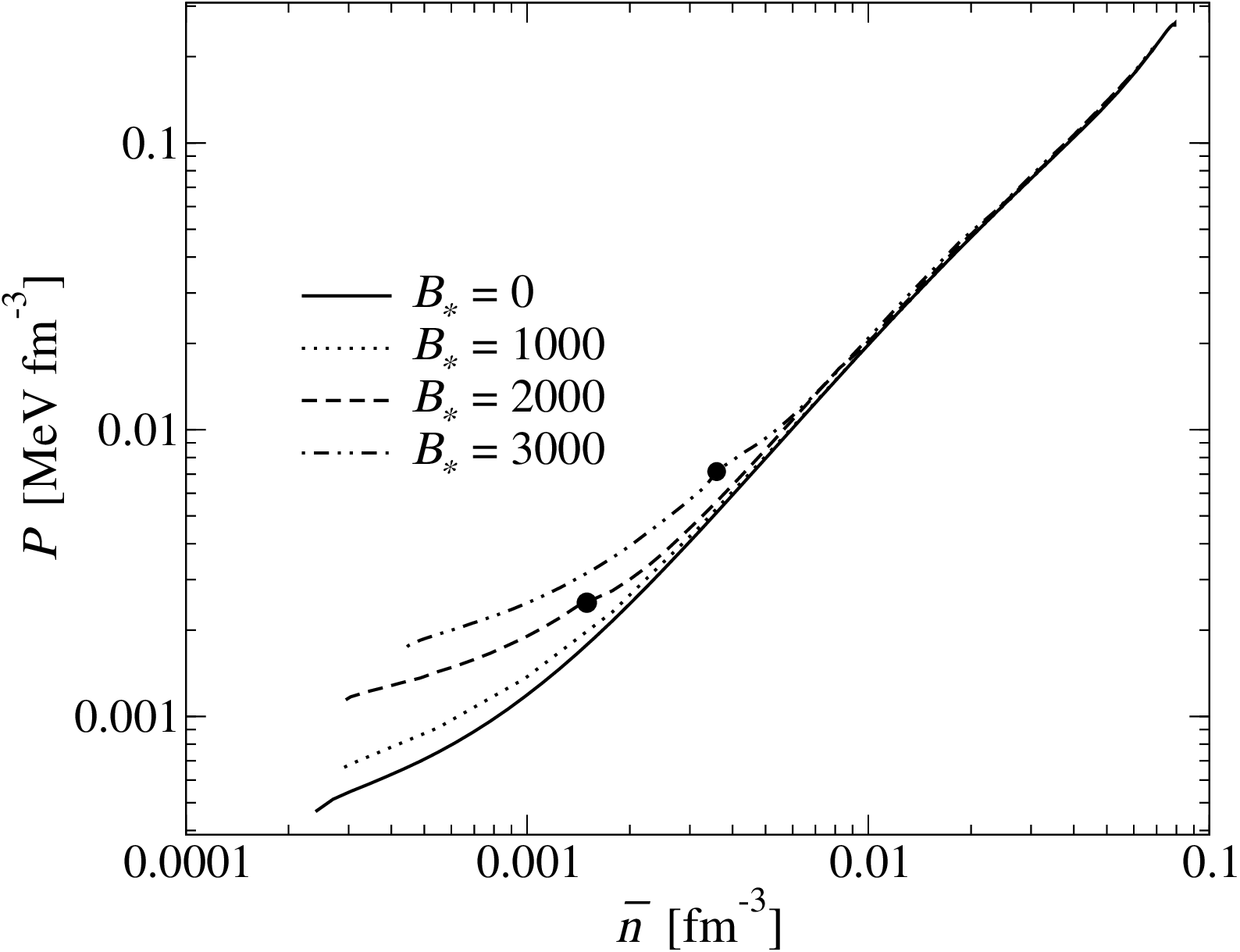}
\caption{Pressure $P$ in MeV~fm$^{-3}$ as a function of the mean baryon number density $\bar n$ in fm$^{-3}$ in the inner crust of a magnetar for different magnetic field strengths $B_\star$. The black circles mark the pressure and density above which the lowest Landau-Rabi level is completely filled.}
\label{fig:P_vs_n-inner}
\end{center}
\end{figure}

The equation of state of the outer and inner regions of the crust is plotted in Fig.~\ref{fig:P_vs_n} for different magnetic field strengths. As previously found in Ref.~\cite{chapav12}, the effects of the magnetic field are most prominent in the outermost region of the crust, and become less and less important with increasing density as more and more Landau-Rabi Levels are filled by electrons. As shown in Fig.~\ref{fig:P_vs_n-inner}, the magnetic field is strongly quantizing in the inner crust (lowest level partially occupied) up to a mean baryon density $\bar{n}\simeq 0.0016$ fm$^{-3}$ and $0.0036$ fm$^{-3}$ for $B_\star=2000$ and $B_\star=3000$ respectively. For $B_\star=1000$, several levels are populated in any region of the inner crust so that the magnetic field has a much smaller impact on the equation of state. In all three cases, the equation of state almost exactly matches that obtained in the absence of magnetic fields at densities above $\bar{n}\sim 0.01$ fm$^{-3}$.

\section{Conclusion}

We have studied the role of Landau-Rabi quantization of electron motion on the equilibrium properties of magnetar crusts, treating consistently both the outer and inner regions in the framework of the nuclear-energy density functional theory. For the outer crust, we have made use of experimental data from the 2016 AME~\cite{ame16} and recent measurements of copper isotopes~\cite{welker2017}, supplemented with the HFB-24 atomic mass model~\cite{gor13}. For the inner crust, we have employed the ETFSI method with the same functional BSk24 as that underlying the HFB-24 model. 

The effects of the magnetic field on the crustal properties are most prominent in the strongly quantizing regime, whereby electrons occupy the lowest Landau-Rabi level only. The magnetic field alters the stratification of the outer crust: the pressure at the interface between two adjacent layers and their densities are shifted; the composition may change as well depending on the magnetic field strength. For $B\sim 10^{17}$~G, the outer crust contain the new nuclides  $^{88}$Sr, $^{126}$Ru, $^{128}$Pd, $^{132}$Sn and $^{124}$Zr, whereas $^{64}$Ni, $^{66}$Ni, $^{78}$Ni,  $^{80}$Ni, $^{80}$Zn and $^{121}$Y are no longer present. The nuclides $^{130}$Cd and $^{79}$Cu that were previously predicted using the 2012 AME~\cite{chapav12,chamut17} have now been ruled out by the latest experimental data. The internal constitution of the inner crust is found be moderately modified. Most regions are still predicted to be made of nuclear clusters with $Z=40$ as in unmagnetized neutron stars~\cite{pearson2018} although other values may be favored depending on the density and on the magnetic field strength. Proton shell and pairing effects, which were neglected in previous investigations~\cite{nandi2011}, are found to play a crucial role for determining the equilibrium composition. Although the shift in $Z$ changes the cluster size, their shape remain essentially unaffected by the magnetic field. Likewise, the magnetic field is not found to have any significant effect on the neutron liquid. 

The magnetic condensation of the outermost layers due to the confinement of electrons in the lowest Landau-Rabi level makes the equation of state very stiff. However, the role of the magnetic field is mitigated by the increasingly larger number of occupied levels in the deeper region of the crust. The equation of state of the inner crust thus remains almost unchanged for magnetic fields below $\sim 10^{16}$~G. For higher magnetic fields of order $10^{17}$~G, the equation of state is moderately modified in the shallowest layers of the inner crust but matches smoothly that of unmagnetized crust at densities above $\sim 0.01$ fm$^{-3}$. 

The properties of magnetar crusts may be further altered due to the effects of the magnetic field on nucleons. These effects, which have been previously studied within the Thomas-Fermi approximation for fixed proton fractions~\cite{lima2013}, may become important for magnetic field strength exceeding $10^{17}$~G. This deserves further studies.

\section*{Acknowledgments}
This work was financially supported by Fonds de la Recherche Scientifique - FNRS (Belgium), NSERC (Canada), and the Bulgarian Academy of Sciences through the program for support of young scientists under contract No.~DFNP-17-167/03.08.2017.
The work of Y. D. M. was also supported by a Short Term Scientific Mission (STSM) grant from the European Cooperation in Science and Technology (COST) Action CA16214. Y. D. M. and Z. S. also acknowledge ``Science Forge'' and MedTech.bg initiative for aspiring scientists.

\end{document}